\documentclass[10pt,conference]{IEEEtran}
\usepackage{dcolumn}
\usepackage{fancyheadings}
\usepackage{amstext,amssymb}
\usepackage{graphicx}
\usepackage{times,color}
\usepackage[final,ulem=normalem]{changes}
\usepackage{subfig}
\usepackage{algorithm}
\usepackage{algorithmic}
\usepackage{latexsym}
\usepackage{amsmath}
\usepackage{flushend}

\newcommand{\comment}[1]{}

\definecolor{dgreen}{rgb}{0,0.48,0.3}

\long\def\comment#1{}

\IEEEoverridecommandlockouts

\definechangesauthor{RW}{blue}
\definechangesauthor{JW}{red}
\definechangesauthor{RW2}{orange}
\definechangesauthor{GD}{magenta}

\begin{document}
\pagestyle{empty}
\title{Mutual-Information Optimized Quantization for LDPC Decoding of Accurately Modeled Flash Data 
\thanks{This research was supported by a gift from Inphi Corp. and UC Discovery Grant 192837.}}


\author{Jiadong Wang, Guiqiang Dong, Tong Zhang and Richard Wesel\\
wjd@ee.ucla.edu,~dongguiqiang@gmail.com,~tzhang@ecse.rpi.edu,~wesel@ee.ucla.edu}

\maketitle \thispagestyle{empty}
\begin{abstract}
High-capacity NAND flash memories use multi-level cells (MLCs) to store multiple bits per cell and achieve high storage densities. Higher densities cause increased raw bit error rates (BERs), which demand powerful error correcting codes. Low-density parity-check (LDPC) codes are a well-known class of capacity-approaching codes in AWGN channels. However, LDPC codes traditionally use soft information while the flash read channel provides only hard information.  Low resolution soft information may be obtained  by performing multiple reads per cell with distinct word-line voltages.  

We select the values of these word-line voltages to maximize the mutual information between the input and output of the equivalent multiple-read channel under any specified noise model. Our results show that maximum mutual-information (MMI) quantization provides better soft information for LDPC decoding given the quantization level than the constant-pdf-ratio quantization approach. We also show that adjusting the LDPC code degree distribution  for the quantized setting provides a significant performance improvement.
\end{abstract}
\section{Introduction}
Flash memory is a low-power non-volatile device that can carry a large amount of data within a small area. The original NAND flash \replaced{architecture}{memories} was called single-level-cell (SLC) flash and used only one nonzero charge level to store one bit. \replaced{More}{The} recent devices use multiple levels and are referred to as multiple-level cell (MLC) flash. Four and eight levels per cell can be found in use, and the number of levels will increase further to provide more storage capability \cite{LiISSCC08}\cite{TrinhISSCC08}.


The increase in the number of levels and aggressive feature-size reduction cause cell-to-cell interference and retention noise to become more severe than for the original SLC flash memories \cite{LeeElectron02}. \replaced{Powerful codes are required}{We need to use stronger codes} to cope with these \replaced{obstacles}{noise} and maximize the potential of the system.


Low-density parity-check (LDPC) codes are a class of capacity-approaching codes for the AWGN channel \cite{RichardsonDes}. Flash systems typically only provide hard reliability information after the reading process while LDPC codes typically utilize soft reliability information. This paper \added{uses realistic channel models to} demonstrate\deleted{s} that efficiently extracting soft information {with} a few extra reads in the cell can significantly improve \deleted{the} LDPC code\deleted{s} performance in flash memory. \deleted{We also showed in \cite{JWANGGLOBECOM11} that just a few more \replaced{cell reads}{soft information} can recover most of the benefit of using the infinite-precision soft information without an unnecessary penalty in reading time.}

Our previous analysis \cite{JWANGGLOBECOM11} use\replaced{d}{s} pulse-amplitude modulation (PAM) with Gaussian noise to model Flash cell threshold voltage levels. We investigate\replaced{d}{s} how to select the word-line voltages to maximize the mutual information between {the} input and {the} output of the equivalent read channel. With carefully selected word-line voltage for each of the reads, we represent the multiple-read channel as a probability transition matrix and decode the data \replaced{using}{with} a standard belief-propagation algorithm. \added[JW]{The maximum mutual information (MMI) approach is also explored in \cite{LeeISIT05} \cite{KurkoskiIT} for the design of the message-passing decoders of LDPC codes to optimize the quantization of the binary-input channel output.}

This paper extends the analysis in \cite{JWANGGLOBECOM11} to any noise model for the flash memory read channel.  As an example, we model a four-level six-read MLC as a four-input seven-output discrete channel. Instead of assuming Gaussian noise distributions as in \cite{JWANGGLOBECOM11}, we numerically compute the probability transition matrix using the retention noise model from \cite{DongUSENIX2011}. 

\begin{figure}
\centering
\includegraphics[width=0.45\textwidth]{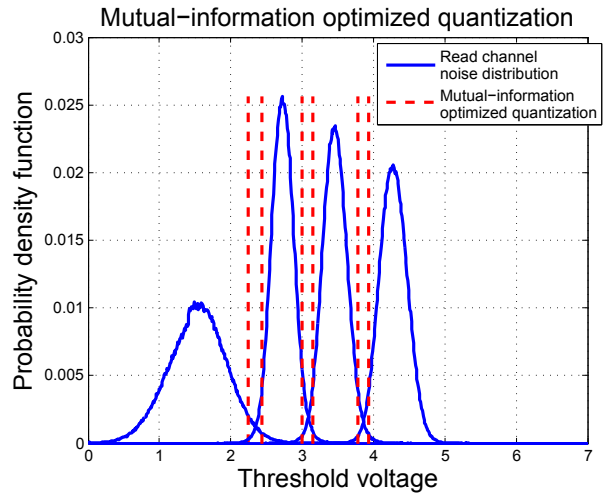}
\caption{Mutual-information optimized quantization for the six-month data.}\label{fig:pdf}
\vspace{-0.25in}
\end{figure}

Fig. \ref{fig:pdf} shows the \added{four conditional threshold-voltage} probability density function\added{s} generated \replaced{according to}{from} \cite{DongUSENIX2011} and the resulting \added{six} MMI word-line voltages after six months retention time.  \added{While the conditional noise for each transmitted (or written) threshold voltage is similar to that of a Gaussian, the variance of the conditional distributions varies greatly across the four possible threshold voltages.  Note that the lowest threshold voltage has by far the largest variance.}

In \cite{DongTCAS2011}, a heuristic quantization algorithm sets the word-line voltages to the value where the two adjacent probability density functions have a constant ratio $ R $. This paper compares the MMI approach with the constant-ratio method \added{of \cite{DongTCAS2011} using the realistic channel model of \cite{DongUSENIX2011} } and shows that \added{the} MMI approach generally outperforms the constant-ratio method.

\added[JW]{This paper also explores how the quantized setting should be considered in the selection of the LDPC degree distribution.  LDPC codes are usually designed with the degree distribution optimized for the AWGN channel \cite{RichardsonDes}. However, our simulations show that, in the quantized setting, adjusting this ``optimal'' degree distribution can significantly improve performance.}

Section~\ref{background} introduces the basics of the NAND flash memory model and LDPC codes.  Section~\ref{casestudy} describes the retention noise model using MLC as an example and shows how to obtain word-line voltages by maximizing the mutual information of the equivalent read channel. Section~\ref{results} provides simulation results demonstrating the benefits adjusting the degree distribution to the quantized setting and compares the MMI approach with the constant ratio method for extracintg soft information. Section~\ref{conclusion} delivers the conclusions.

\section{Background}\label{background}
This section introduces the basics of NAND flash memory and LDPC codes.

\subsection{Basics of NAND Flash Memory}\label{basic:NAND}
This paper focuses on the NAND-architecture flash memory that is currently the most prevalent architecture. In the NAND architecture, each memory cell features a transistor with a control gate and a floating gate. To store information, a relatively large voltage is applied to the control gate, which adds a specified amount of charge to the floating gate through Fowler-Nordheim tunneling  \cite{BezIEEE03}.

\begin{figure}
\centering
\includegraphics[width=0.4\textwidth]{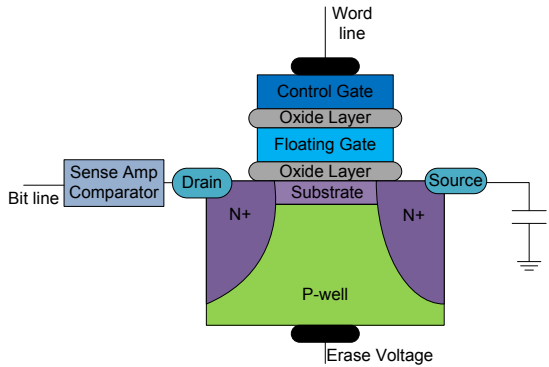}
\caption{A NAND flash memory cell.}\label{flashcell}
\end{figure}

Fig. \ref{flashcell} shows the configuration of a NAND flash memory cell. To read a memory cell, the charge level written to the floating gate is detected by applying a specified word-line voltage to the control gate and measuring the transistor drain current. The drain current is compared to a threshold by a sense amp comparator. If the drain current is above the comparator threshold, then the word-line voltage was sufficient to turn on the transistor, indicating that the charge written to the floating gate was insufficient to prevent the transistor from turning on. If the drain current is below the threshold, the charge added to the floating gate was sufficient to prevent the applied word-line voltage from turning on the transistor. 

The sense amp comparator only provides one bit of information about the charge level present in the floating gate. The word-line voltage required to turn on a particular transistor (called the threshold voltage) can vary from cell to cell for a variety of reasons. For example, the floating gate can receive extra charge when nearby cells are written, the floating gate can be overcharged during the write operation, or the floating gate can lose charge due to leakage in the retention period \cite{MaedaISDFT09}. We are not optimizing word-line voltages for a particular cell, but rather for the threshold voltage distribution over all cells at a certain retention time.

In \cite{JWANGGLOBECOM11}, we assumed an i.i.d. Gaussian threshold voltage for each level of an MLC flash memory cell.  More precise models such as the model in \cite{MaedaISDFT09} in which the lowest and highest threshold voltage distributions have a higher variance and the model in \cite{LiVLSI10} in which the lowest threshold voltage (the one associated with zero charge level) is Gaussian and the other threshold voltages have Gaussian tails but a uniform central region are sometimes used. The model in  \cite{DongUSENIX2011} is similar to \cite{LiVLSI10}, but is derived by explicitly accounting for \replaced[GD]{real dominating noise sources, such as inter-cell interference, program injection statistics, random telegraph noise and retention noise}{ inter-cell interference}.  \deleted{Despite its limitations, the simple Gaussian model is sufficient to motivate the proposed investigation of soft information.  Furthermore, the techniques presented in this paper can easily be extended to other probability distributions.}  \added{In this paper, we use the read channel model of \cite{DongUSENIX2011}.}


\subsection{Basics of LDPC codes} \label{basic:LDPC}
LDPC codes are well-known as capacity-approaching codes of the AWGN channel, and they are defined by sparse parity-check matrices. The degree distribution of LDPC codes can be optimized to operate closely to the capacity of an AWGN channel \cite{RichardsonDes}. For a given degree distribution, we can generate LDPC codes using several algorithms, such as the PEG algorithm \cite{PEG01}, and the ACE algorithm \cite{TianTCOM04}.

Storage systems typically require frame-error-rates lower than $ 10^{-15} $, making the design of LDPC codes with low error-floors necessary for applications to flash memory. This topic has generated a significant amount of recent research including \cite{richardson} \cite{JWANGICC11} \cite{ivkovicIT08} \cite{NguyenITW10} \cite{HuangITA2011}.

Iterative belief propagation algorithms are used for decoding LDPC codes.  \deleted{These iterative decoding algorithms are called belief-propagation algorithms.} Soft reliability information at the receiver usually can significantly improve the performance of belief-propagation decoders. Conversely, a coarse quantization of the received information can degrade the performance of an LDPC code.

The remainder of this paper \replaced{presents}{studies} a general quantization approach to select word-line voltages to maximize the mutual information for an $ N $-level $ M $-read Flash memory and any noise model. \added{These word-line voltages are then used to gather quantized soft information for an LDPC decoder.}  We also compare the MMI approach with an alternative quantization approach and show how adjusting the degree distribution can improve performance in a quantized setting.

\section{Quantization for MLC Flash}\label{casestudy}
For any $ N $-level $ M $-read Flash memory and any noise model, the multiple-read channel can be represented by a probability transition matrix after choosing the word-line voltage for each of the reads, and the data can be decoded with a standard belief-propagation algorithm. \replaced{The equivalent discrete channel induces a mutual information between the $ N $ inputs and $ M+1 $ outputs.  

\subsection{Maximum Mutual Information (MMI) Quantization}
The approach is to select the set of word-line voltages, which maximizes that mutual information. } {We can numerically compute the mutual information between the $ N $ inputs and $ M+1 $ outputs of the equivalent discrete channel, and select the set of word-line voltages that maximizes the mutual information.}

\begin{figure}
\centering
\includegraphics[width=0.4\textwidth]{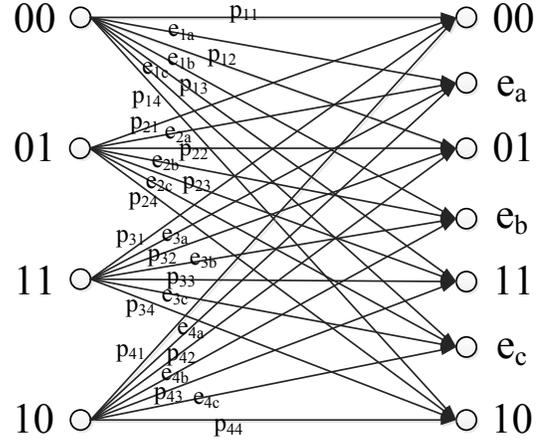}
\caption{Quantization model for 4-MLC with 6 reads.}\label{ch4MLC6reads}
\end{figure}

This section uses a 4-level 6-read MLC with retention noise as an example. For 4-level MLC flash memory, each cell can store 2 bits of information. \replaced{Gray labeling $ (00,01,11,10) $ minimizes the raw bit error rate for these four levels.}{To minimize the raw bit error rate, we use the Gray labeling $ (00,01,11,10) $ for these four levels.} In 4-level MLC flash, each cell is typically compared to three word-line voltages and thus the output of the comparator has four distinct quantization regions. In this section, we consider three additional word-line voltages (for a total of six) and quantize the threshold voltage to seven distinct regions as shown in Fig.~\ref{fig:pdf}.  \replaced{Fig.~\ref{ch4MLC6reads} presents the post-quantization channel model, a 4-input 7-output discrete memoryless channel.}{ An equivalent channel model is given in Fig.~\ref{ch4MLC6reads}. and models it as a 4-input 7-output discrete channel as shown in Fig.~\ref{ch4MLC6reads}. }

Suppose the 6 word-line voltages are $ q_1,q_2,...,q_6 $. We can numerically compute the probability transition matrix using the probability density function generated from the retention noise model in \cite{DongUSENIX2011}. Fig.~\ref{fig:pdf} shows the probability density function generated from \cite{DongUSENIX2011} and the resulting MMI word-line voltages after 6 months retention time.

Since the retention noise model itself is not an analytic expression and certainly not symmetric, we need to numerically compute all the probabilities in Fig.~\ref{ch4MLC6reads} and calculate the mutual information between the input and output:

\small
\begin{align}
&I(X;Y)\notag\\
=& H(Y)-H(Y|X)\notag\\
=& H\left(\frac{p_{11}+p_{21}+p_{31}+p_{41}}{4},\frac{p_{12}+p_{22}+p_{32}+p_{42}}{4}, \right. \notag\\
&\frac{p_{13}+p_{23}+p_{33}+p_{43}}{4},\frac{p_{14}+p_{24}+p_{34}+p_{44}}{4},\notag\\
&\frac{e_{1a}+e_{2a}+e_{3a}+e_{4a}}{4},\frac{e_{1b}+e_{2b}+e_{3b}+e_{4b}}{4},\notag\\
&\left. \frac{e_{1c}+e_{2c}+e_{3c}+e_{4c}}{4}\right)\notag\\
&-\frac{1}{4}H(p_{11},p_{12},p_{13},p_{14},e_{1a},e_{1b},e_{1c})\notag\\
&-\frac{1}{4}H(p_{21},p_{22},p_{23},p_{24},e_{2a},e_{2b},e_{2c})\notag\\
&-\frac{1}{4}H(p_{31},p_{32},p_{33},p_{34},e_{3a},e_{3b},e_{3c})\notag\\
&-\frac{1}{4}H(p_{41},p_{42},p_{43},p_{44},e_{4a},e_{4b},e_{4c}).\label{equ:mu6reads_asym}
\end{align}
\normalsize

\added[JW]{The mutual information in \eqref{equ:mu6reads_asym} is in general not a quasi-concave function in terms of the word-line voltages $ q_1,q_2,...,q_6 $, although it is quasi-concave for the simple model of two symmetric Gaussians with symmetric word-line voltages studied in \cite{JWANGGLOBECOM11}. Since \eqref{equ:mu6reads_asym} is a continuous and smooth function and locally quasi-concave in the range of our interest, we can numerically compute the maximum mutual information with a careful use of bisection search.}

After optimizing the word-line voltages $ q_1,q_2,...,q_6 $, the equivalent 4-input 7-output discrete channel has the maximum mutual information between the input and output. We can easily extend this technique to any other $ N $-level $ M $-read Flash memory and any other noise model with known conditional distributions.

\subsection{Constant PDF-Ratio Quantization}
In \cite{DongTCAS2011},  a \deleted{heuristic} quantization algorithm sets \replaced{each}{the} word-line voltage\deleted{s} to the value where the two adjacent pdfs have a constant ratio $ R $.  The difficult step in this algorithm is the selection of $R$, which is accomplished by a heuristic process of simulation and adjustment.  The goal of simulation and adjustment is to optimize the decoder performance.

\begin{figure} [b]
\centering
\includegraphics[width=0.4\textwidth]{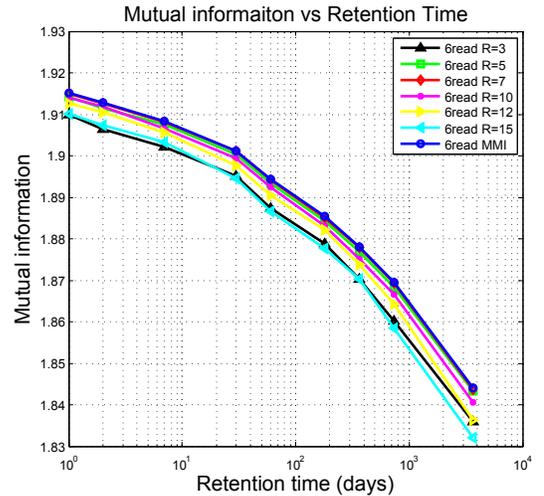}
\caption{Mutual information comparison between the MMI approach and constant ratio method with retention data.}\label{fig:MIcomp}
\end{figure}

Fig.~\ref{fig:MIcomp} compares the mutual information obtained by using the MMI approach and the constant ratio method for a variety of $R$ values.   We note that with certain $R$ values, the constant ratio method can provide mutual information that is close to the maximum.  However, finding the best $R$ can be a challenge.

\section{LDPC Peformance Comparisons}\label{results}
This paper uses rate-0.9021 irregular LDPC codes with block length $ n=9118 $ and dimension $ k=8225 $ for \deleted{the} simulation\deleted{s}. LDPC matrices are constructed according to the degree distributions using the ACE algorithm \cite{TianTCOM04} together with the stopping-set check algorithm \cite{RamamoorthyICC04} to optimize the LDPC matrix. Simulations were performed using a sequential belief propagation decoder (layered belief propagation) \cite{YEOGLOBECOM01}. \added[JW]{A rate-0.9021 BCH code with block length $n=9152$ and dimension $k=8256$ provides a baseline for comparison.}

\subsection{Degree distribution in a quantized setting}
Two of the LDPC codes studied feature distinct degree distributions with maximum variable degree 19. For Code 1, the degree distribution is the usual optimal degree distribution for AWGN \cite{RichardsonDes}.  For Code 2, the initial AWGN-optimal degree distribution is adjusted to improve performance in a quantized setting as follows:

\added[JW]{Hard decoding induces small absorbing sets such as $ (4,2)$, $(5,1)$, $(5,2) $ absorbing sets in Code 1.  To preclude these absorbing sets, we increase all the degree-3 variable nodes to have degree 4 to produce Code 2. Code 2 significantly outperforms Code 1 under hard decoding. 

\begin{figure}[t]
\centering
\includegraphics[width=0.45\textwidth]{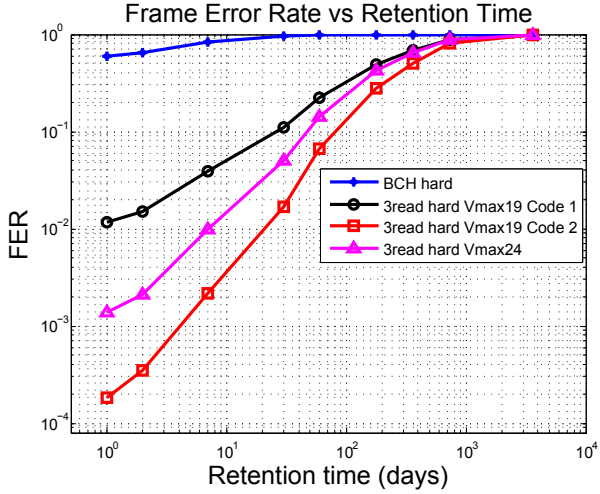}
\caption{Simulation results for 4-level MLC using hard quantization.}\label{fig:simcodecompare}
\end{figure}

\added[JW]{We also simulated another code with the maximum variable degree 24 with degree distribution optimized for AWGN (Code 3). Fig.~\ref{fig:simcodecompare} shows frame error rate versus retention time under hard decoding for these three codes.} 
Code 3 has an even better threshold in AWGN than Code 1, but the newly designed Code 2 with the lower AWGN threshold still outperforms it under hard quantization. This demonstrates that a superior AWGN threshold does not necessarily imply superior performance under hard decoding.  Of course when simulated in AWGN with full resolution soft decoding, Code 3 performs better than Code 1, and Code 1 performs better than Code 2 .}   

\subsection{Comparison of quantization methods}
\begin{figure}
\centering
\includegraphics[width=0.45\textwidth]{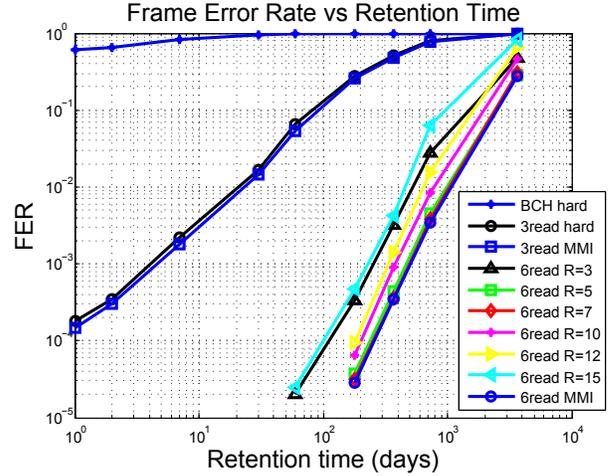}
\caption{Simulation results for 4-level MLC using MMI and constant R with retention data and Code 2.}\label{fig:simretention_code2}
\vspace{-0.1in}
\end{figure}

Fig.~\ref{fig:simretention_code2} shows frame error rate (FER) plotted versus retention time for Code 2 under a variety of quantizations. This LDPC code has the adjusted degree distribution to lower the FER for hard quantization.  Since the noise model is not symmetric, the MMI approach with 3 reads has a slightly larger mutual information than the hard quantization with 3 reads, and thus performs slightly better. 

\begin{figure}
\centering
\includegraphics[width=0.45\textwidth]{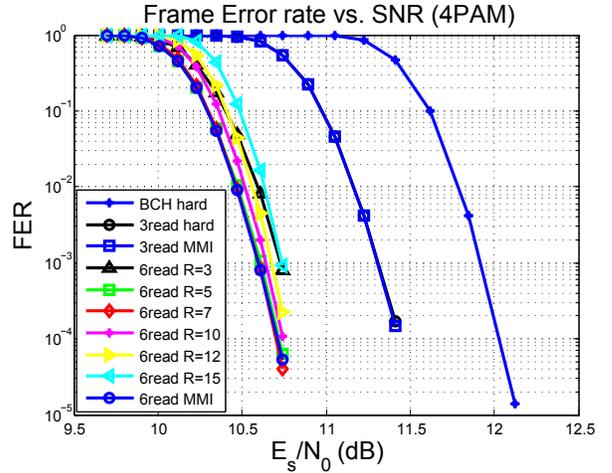}
\caption{Simulation results for 4-level MLC using MMI and constant R with the Gaussian model and Code 2.}\label{fig:simgaussian_code2}
\vspace{-0.2in}
\end{figure}

In this plot, the performance of the LDPC codes is closely related to the mutual information of the equivalent channel given in Fig. \ref{fig:MIcomp}. The $ R=7 $ case has the closest mutual information to the MMI and its performance is also very close to that of the MMI approach. The MMI approach outperforms the constant $ R $ method with most $ R $s in this plot. For comparison, Fig.~\ref{fig:simgaussian_code2} shows similar results for six reads using the Gaussian model in \cite{JWANGGLOBECOM11}.

\begin{figure}
\centering
\includegraphics[width=0.45\textwidth]{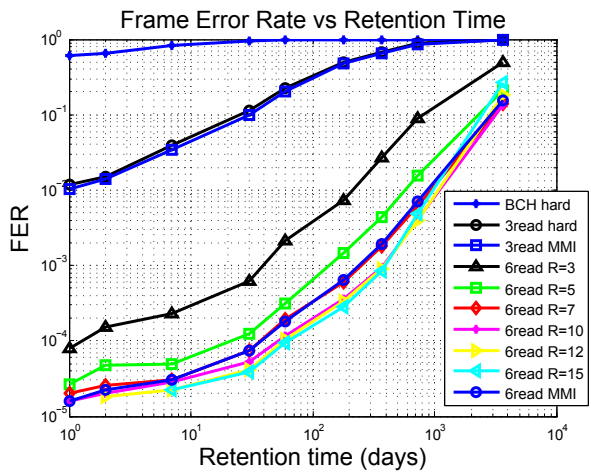}
\caption{Simulation results for 4-level MLC using MMI and constant R with retention data and Code 1.}\label{fig:simretention}
\vspace{-0.1in}
\end{figure}

\begin{figure}
\centering
\includegraphics[width=0.45\textwidth]{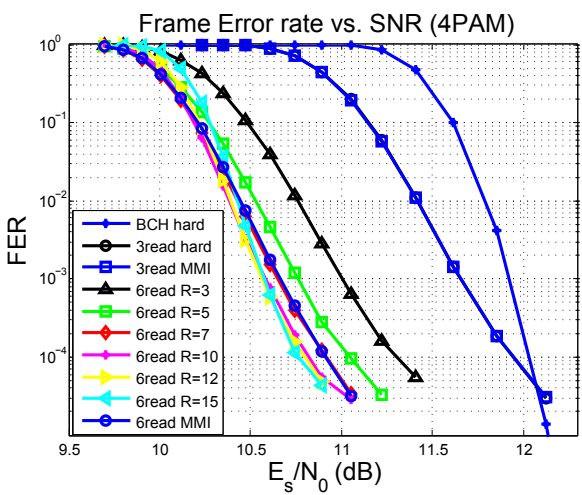}
\caption{Simulation results for 4-level MLC using MMI and constant R with the Gaussian model and Code 1.}\label{fig:simgaussian}
\vspace{-0.2in}
\end{figure}

Figs. \ref{fig:simretention} and \ref{fig:simgaussian} are analogous to Figs. \ref{fig:simretention_code2}  and \ref{fig:simgaussian_code2} but for Code 1, which was not adjusted for the quantized setting.   In this case the constant ratio method with $ R=15 $ slightly outperforms the MMI approach. For comparison, Fig.~\ref{fig:simgaussian} shows similar results for six reads using the Gaussian model.  This example reflects our experience that the only cases in which the constant ratio method (slightly) outperforms the MMI method are cases in which the LDPC degree distribution is not well-matched to the channel.  In other words, when one has identified a good code for the channel, MMI will give the best quantization.  The degree distribution for which MMI was not optimal was also not the best choice of degree distribution.

\section{Conclusion}\label{conclusion}
This paper shows that using a small amount of soft information significantly improves the performance of LDPC codes and demonstrates a clear performance advantage over conventional BCH codes. In order to maximize the performance benefit of the soft information, we develop a word-line-voltage-selection method that maximizes the mutual information between the input and output of the \added{DMC} equivalent \added{to the quantized} read channel. This method can be applied to any given channel model\deleted{s} and provides an effective and efficient estimate of the word-line voltages\added{, as} compared to other existing quantization techniques. Possible directions for future research include the design of better high-rate LDPC codes specifically for the flash memory channel, and the analysis of the corresponding error-floor properties.




\bibliographystyle{unsrt}	
\bibliography{myrefs}		

\end{document}